\documentclass{eas}
\usepackage{graphicx}
\usepackage{amssymb}

\TitreGlobal{The Title of this Volume}
\begin{document}

\title{Three Dimensional Molecular Line Transfer Study of
  Extragalactic ISM : AGN/Starburst Connection} 
\author{Masako YAMADA}\address{ALMA Project Office, National
  Astronomical Observatory of Japan}
\runningtitle{M. Yamada : Line Transfer Study of Extragalactic ISM}
\begin{abstract}
Molecular gas in external galaxies is a subject of crucial importance
for observational and theoretical studies of galaxy formation.
Compact molecular gas around an active galactic nuclei (AGN) is
expected to be an energy budget of AGN and/or the nuclear
starburst.
Recent observational studies suggest that line ratios in
millimeter and submillimeter band may be a good tool to reveal the
long-standing question on the origin of activity -- AGN or nuclear
starburst. 
We have constructed a powerful ``telescope'' of theory,
three-dimensional nonLTE line transfer code, preceding the high
resolution and sensitivity observations such as ALMA.
\end{abstract}
\maketitle
\section{Introduction}
Current progress in millimeter and submillimeter observations have
found peculiar signature in ratios of molecular lines towards numbers
of compact gas at the centers of active galaxies.
The pioneering works of  Kohno {\em et al.} (\cite{kohno2001,
  kohno2005}) showed that some of
nearby Seyfert galaxies have enhanced ratio of
$R_{\mathrm{HCN/HCO}^{+}}$ as high as 2.5 or more.
Following observations of various molecular lines have given rise of
discussion of peculiar chemistry
such as photon dominated region (PDR) or X-ray dominated region (XDR)
(Meijerink {\em et al.} \cite{meijerink2005, meijerink2007}, Aalto
{\em et al.} \cite{aalto2007}.)
However, since the compact molecular core is not resolvable with
current mm/submm observational instruments, one has to take into
account of internal structures within a core.
Hydrodynamic simulations present highly inhomogeneous density and
temperature structure within a compact molecular gas at the center of
a galaxy (Wada \& Tomisaka \cite{wada2005}).
In an inhomogeneous molecular gas naturally expected is the
inhomogeneous population distribution. 
We performed three-dimensional non-thermodynamic equilibrium (nonLTE)
molecular line transfer simulations, and examined the effects of
excitation status inside a clumpy torus.

\section{Ratio of HCN and HCO$^{+}$ Lines in Millimeter Band}

\subsection{Simulations}

We first performed hydrodynamic simulations of molecular gas in a
steady gravitational potential of supermassive black hole of
$M_\mathrm{SMBH} = 10^8M_{\odot}$ and galactic halo (see for details,
Wada \& Tomisaka \cite{wada2005}, Yamada {\em et al.},
\cite{yamada2007b}). 
The hydrodynamic simulation explicitly includes radiative cooling,
feedback from supernovae, strong UV heating ($G_0=10$) for thermal
evolution as well.
Resultant molecular ISM has a highly inhomogeneous structure in
density, temperature, and turbulent velocity. 
Next we calculated three-dimensional nonLTE line transfer of HCN and
HCO$^{+}$ pure rotational lines using a snapshot data of hydrodynamic
simulation.
We assumed spatially uniform molecular abundance distribution in order
to examine the effect of clumpy excitation conditions on the line
ratio $R_{\mathrm{HCN/HCO}^+}\equiv
I_{\mathrm{HCN(1-0)}}/I_{\mathrm{HCO}^{+}(1-0)}$ separately from
chemistry models.

\subsection{Intensity Distribution and $R_{\mathrm{HCN/HCO}^{+}}$}

\begin{center}
\begin{figure}[thbp]
  \includegraphics[width=10cm]{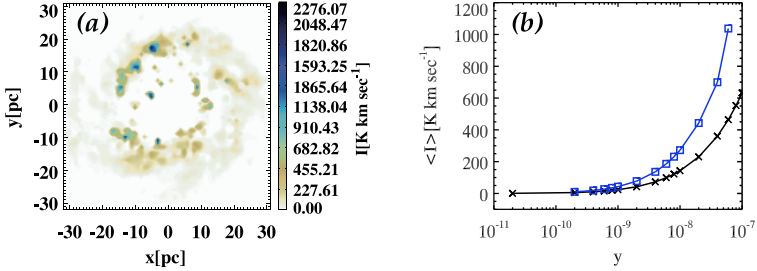}
  \caption{Integrated intensity distribution of HCN(1-0) line of line
  transfer simulation with $y=2\times 10^{-9}$ (left panel).  Right
  panel shows integrated intensity averaged over the field-of-view as
  a function of $y$. Line with squares denotes HCO$^{+}$(1-0) line,
  and line with crosses denots HCN(1-0) line, respectively.}
  \label{fig:result1}
\end{figure}
\end{center}
In Figure \ref{fig:result1} (a) we present integrated intensity
distribution of HCN(1-0) ($I_\mathrm{HCN}$) line. 
The distribution of $I_\mathrm{HCN}$ shows clumpy structure,
reflecting the highly inhomogeneous internal structure of the model
torus.
The typical size of the bright spots in Fig. \ref{fig:result1} is
$\approx 0.03^{\prime\prime}$ if the torus is located at the distance
of $D=$ 20 Mpc. 
Thus though current mm-telescopes smear out these internal structures,
high angular resolution of forthcoming telescope
such as Atacama Large Millimeter/submillimeter Array (ALMA) will
reveal inhomogeneous structures in molecular tori in
distant galaxies. 
The Fig.\ref{fig:result1} (b) plots integrated intensities of HCN(1-0)
and HCO$^{+}$(1-0) avraged over the ``field of view'' of our
simulations as a function of fractional molecular abundance $y$.
For the same value of $y$, $\langle I_{\mathrm{HCO}^{+}} \rangle$ is
always larger than $\langle I_\mathrm{HCN} \rangle$, which means that  
a ratio of $R_{\mathrm{HCN/HCO}^{+}}\sim 2$ observed in some of nearby
galaxies requires onverabundant HCN compared with HCO$^{+}$
($y_\mathrm{HCN} \approx 10\times y_{\mathrm{HCO}^{+}}$).
This result and recent chemical studies of PDR and XDR do not seem to
support the original idea of mm/submm molecualr line diagnostics of
AGN/starbust connection (e.g. Kohno et al. \cite{kohno2001}).
Other models, such as mid-InfraRed (MIR) photon pumping
(e.g., Graci\'{a}-Carpio \etal \cite{gracia-carpio2006}; Aalto \etal
\cite{aalto2007}) or hot core chemistry (e.g., St\"{a}uber \etal
\cite{stauber2005}) have been proposed, though, theoretical model
description of molecular line ratios been controversial.

\section{Multi-Transition Analysis in Inhomogeneous Torus}

\begin{figure}[htbp]
  \includegraphics[width=9cm]{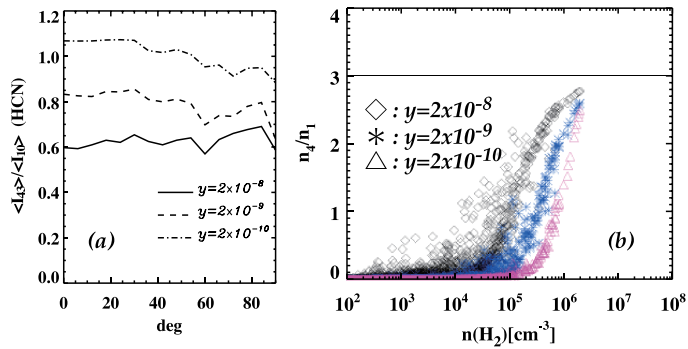}
  \caption{$R_{43/10}$ as a function of viewing angle with respect to
  rotational axis of the torus. 
  Right panel shows population ratio of $J=4$ and 1, or excitation
  temperature of $T_\mathrm{ex}(41)$.}
  \label{fig:result2}
\end{figure}

Current progress in submm observations enables multi-transion analysis
and probe of warm and dense gas at the centers of external galaxies.
In Figure \ref{fig:result2} ratio of integrated intensities $\langle
I_{\mathrm{HCN}(4-3)} \rangle$ and $\langle I_{\mathrm{HCN}(1-0)}\rangle$
(hereafter denoted as $R_{43/10}$) is plotted as a function of
fractional molecular abundance $y$.
In our simulation results ratio $R_{43/10}$ clearly decreases with $y$.
%
%

On the other hand, one-zone analysis predicts a reverse trend : if $y$
is small, optical thickness $\tau_0$ is expected to be small, then the
ratio $R_{43/10}$ will be
\begin{equation}
  R_{43/10}\approx
  \frac{h\nu_{43}n_4A_{43}/(h\nu_{43})^2}{h\nu_{10}n_1A_{10}/(h\nu_{10})^2} 
  \propto \frac{\sum_{J^{\prime}\ge 4}\gamma_{0J^{\prime}}}{\sum_{J\ge
  1}\gamma_{0J}}
  \approx 0.3, 
\end{equation}
where $\gamma_{0J}$ is collisional excitation constant from energy
level $J=0$ to $J$. 
If $y$ is large, $\tau_0$ will be also large, and $R_{43/10}$
approaches unity.

This result indicates the importance of the inhomogeneity in structure
of the torus. 
We model the clumpy torus as two-phase ISM consists of 1) dense clumps
with $n\sim n_\mathrm{crit}$ and 2) tenuous ambient medium with $n\ll
n_\mathrm{crit}$. 
If we assume high kinetic temperature ($k_BT_\mathrm{kin}\gg h\nu$)
and optically thin over a whole region, intensity ratios become (see
eqs. [9] and [10] of Yamada et al. \cite{yamada2007a})
\begin{eqnarray}
  \langle R^{\prime}_{43/10} \rangle_d &\simeq& 
     \frac{\langle n_4A_{43}h\nu_{43}\rangle}{\langle n_4A_{43}h\nu_{43}\rangle} 
     \sim \frac{4^5}{3}\left( \frac{n_4}{n_1}\right) \lesssim 1000,  \\
  \langle R^{\prime}_{43/10} \rangle_t &\simeq&
     \frac{\langle h\nu_{43}\sum_{J\ge 4} \gamma_{0J}n_0\rangle}{\langle h\nu_{10}\sum_{J\ge 1} \gamma_{0J} n_0\rangle}
     \sim J = 4.
\end{eqnarray}
Mean ratio $\langle R_{43/10}\rangle_\mathrm{all}$ in a synthetic
observation beam can be written as a kind of average of $\langle
R_{43/10} \rangle_d$ and $\langle R_{43/10} \rangle_t$,
\begin{equation}
  \langle R^{\prime}_{43/10}\rangle_\mathrm{all}
  = \eta \langle R^{\prime}_{43/10} \rangle_d
    +(1-\eta) \langle R^{\prime}_{43/10} \rangle_t,
  \label{eq:ave}
\end{equation}
where $\eta$ denotes a volume filling factor of dense clumps.
Equation (\ref{eq:ave}) means that if $\eta$ is sufficiently smaller
than unity, $R_{43/10}$ can take a value of order unity {\em even if
  the medium is optically thin.}
In our radiative transfer simulations, ``dense clumps'' can safely be
defined as regions with $n\ge 10^5$ cm$^{-3}$ (close to
$n_\mathrm{crit}$ of HCN(1-0) transition), and $\eta_\mathrm{sim}$ is
indeed $\approx 10^{-3}$.
Furthermore, nonLTE population (occasional $\tau_0 <0$ for $J=$1-0
transitions) and inhomogeneous optical thickness
distribution complicate the $R_{43/10}$ distributions (see
right panel of Fig. \ref{fig:result2}).
Our analysis demonstrates the importance of inhomogeneity of
molecular gas in estimation of physical properties from line ratios
from unresolved observational data.
Inherent inhomogeneity should be paid much attention when interpreting
line ratio in terms of chemical evolution of molecular gas.

\section{Summary}
We examine molecular line diagnostics of molecular gas in the center
of external galaxies using three dimensional radiative transfer
simulations.
Our results demonstrate that realistic inhomogeneous molecular gas
can produce line ratio values unexpected from one-zone analysis, and
emphasize the ability and applicability of our approach, that is,
compilation of high-resolution hydrodynamic simulations and
three-dimensional radiative transfer calculations (``telescope of
theory'') towards forthcoming ALMA era.



\end{document}